\documentclass[conference]{IEEEtran}
\IEEEoverridecommandlockouts
\usepackage{cite}
\usepackage{amsmath,amssymb,amsfonts}
\usepackage{graphicx}
\usepackage{textcomp}
\usepackage[hyphens]{url}
\usepackage{xcolor}

\usepackage[colorlinks=true,
            linkcolor=blue,
            citecolor=blue,
            urlcolor=blue]{hyperref}

\newcommand{\figref}[1]{Fig.~\ref{#1}}

\newcommand*{\expertprompting}{ExpertPrompting}

\begin{document}

\title{Migrating Existing Container Workload to Kubernetes - LLM Based Approach and Evaluation}

\author{\IEEEauthorblockN{Masaru Ueno}
	\IEEEauthorblockA{
		\textit{Fujitsu Limited}\\
		Kawasaki, Japan \\
		ueno.masaru@fujitsu.com}
	\and
	\IEEEauthorblockN{Tetsuya Uchiumi}
	\IEEEauthorblockA{
		\textit{Fujitsu Limited}\\
		Kawasaki, Japan \\
		uchiumi.tetsuya@fujitsu.com}
}

\maketitle

\begin{abstract}
	Although Kubernetes has become a widespread open-source system that automates the management of containerized applications,
its complexity can be a significant barrier, particularly for application developers unfamiliar with it.
One approach employs large language models (LLMs) to assist developers in generating Kubernetes manifests;
however it is currently impossible to determine whether the output satisfies given specifications and is comprehensible.
In this study, we proposed a benchmarking method for evaluating the effectiveness of LLMs in synthesizing manifests,
using the Compose specification --- a standard widely adopted by application developers --- as input.
The proposed benchmarking method revealed that LLMs generally produce accurate results that compensate for simple specification gaps.
However, we also observed that inline comments for readability were often omitted,
and completion accuracy was low for atypical inputs with unclear intentions.

\end{abstract}

\begin{IEEEkeywords}
	Configuration management, Kubernetes, program synthesis, LLM
\end{IEEEkeywords}

\section{Introduction}

Recently, containerization technology has rapidly proliferated within enterprises in the field of software development.
Notably, both Kubernetes~\cite{kubernetes} and Docker Compose~\cite{dockercompose} have been widely used in various environments.
The significant differences in functionality, abstraction level, and configuration complexity between Kubernetes and Compose can make the transition from development to production environments challenging.
Therefore, it is crucial to appropriately address and overcome these gaps.

To address this gap, two existing approaches have been employed: Kompose~\cite{Kompose} and large language models (LLMs).
Kompose converts Docker Compose configuration files into Kubernetes configuration files or \textit{manifests}.
However,
Kompose provides little help with migration
because its output must be manually modified by a proficient Kubernetes expert.

In contrast, LLMs such as ChatGPT~\cite{chatgpt} can assist in the transition from Compose to Kubernetes without requiring developers to have a deep knowledge of Kubernetes.
However, LLMs are still being developed and are not sufficiently reliable.
Previous studies evaluated
the quality of manifests generated by ChatGPT using static analysis tools;
however, these evaluations failed to ensure successful migration because they did not account for correspondence with input specifications or evaluations for developers.

To address the limitations of previous studies,
we developed a new benchmark to quantitatively evaluate the quality of the manifests generated by LLMs.

Our evaluation method examines the following aspects of LLM-generated manifests:

\begin{enumerate}
	\item     Evaluation for developers: that is, whether the generated code is written in an easily understandable format and whether it produces the expected results.
	\item     Correspondence with input specifications: that is, how well the LLM-generated code matches the developer's problem description input.
	\item     Output consistency.
\end{enumerate}

The contributions of the present study are as follows:
\begin{enumerate}
	\item We assessed the limitations of Kompose and demonstrated practical examples of conversion failure.
	\item We developed a benchmark for LLM-generated manifests, focusing on input-adherence and maintainability.
	\item We identified the characteristics and challenges of LLM-based manifest synthesis and suggested directions for future research.
\end{enumerate}

\section{Background}
\subsection{Container Management with Kubernetes}

Kubernetes~\cite{kubernetes} has become the industry standard for container orchestration owing to its declarative approach, comprehensive feature set for efficient managing of complex application environments, and robust community support.
This open-source platform automates the management of containerized applications, making it indispensable for modern software development and deployment.

Kubernetes utilizes manifests, YAML or JSON files, to describe the desired states of the objects in Kubernetes.
\figref{kompose-in-out}(a) shows a sample manifest with four main fields:
\texttt{apiVersion}, \texttt{kind}, \texttt{metadata}, and \texttt{spec}, specifying the API version, resource type, metadata, and object specifications, respectively.
For instance, a Service object specifications include the ports to which the service listens and a label selector to identify the Pods (or containers) that belong to the Service,
whereas the Deployment object specification includes container startup options and volume information.
In particular, creating Service or PersistentVolumeClaim objects generally requires in-depth knowledge of Kubernetes owing to cluster-specific requirements. This complexity presents a significant barrier for application developers unfamiliar with Kubernetes.

\subsection{Container Management with Compose}

Docker Compose is an official Docker tool that simplifies the management of multiple containers, and is widely used in the development of containerized applications.

\figref{kompose-in-out}(b) shows an example of Compose specifications.
Compose specifications are in the YAML format to define and run multi-container applications.
It offers a higher level of abstraction with a more intuitive syntax than Kubernetes for managing containerized services,
allowing users to define services, networks, and data volumes easily.

\subsection{Possible Solutions for Migrating Compose to Kubernetes}

\subsubsection{Using Kompose}
Kompose is currently the primary and almost the sole viable tool for converting Compose specifications into Kubernetes manifests,
which is a Kubernetes project.

\begin{figure}
	\centering
	\includegraphics[width=.77\columnwidth]{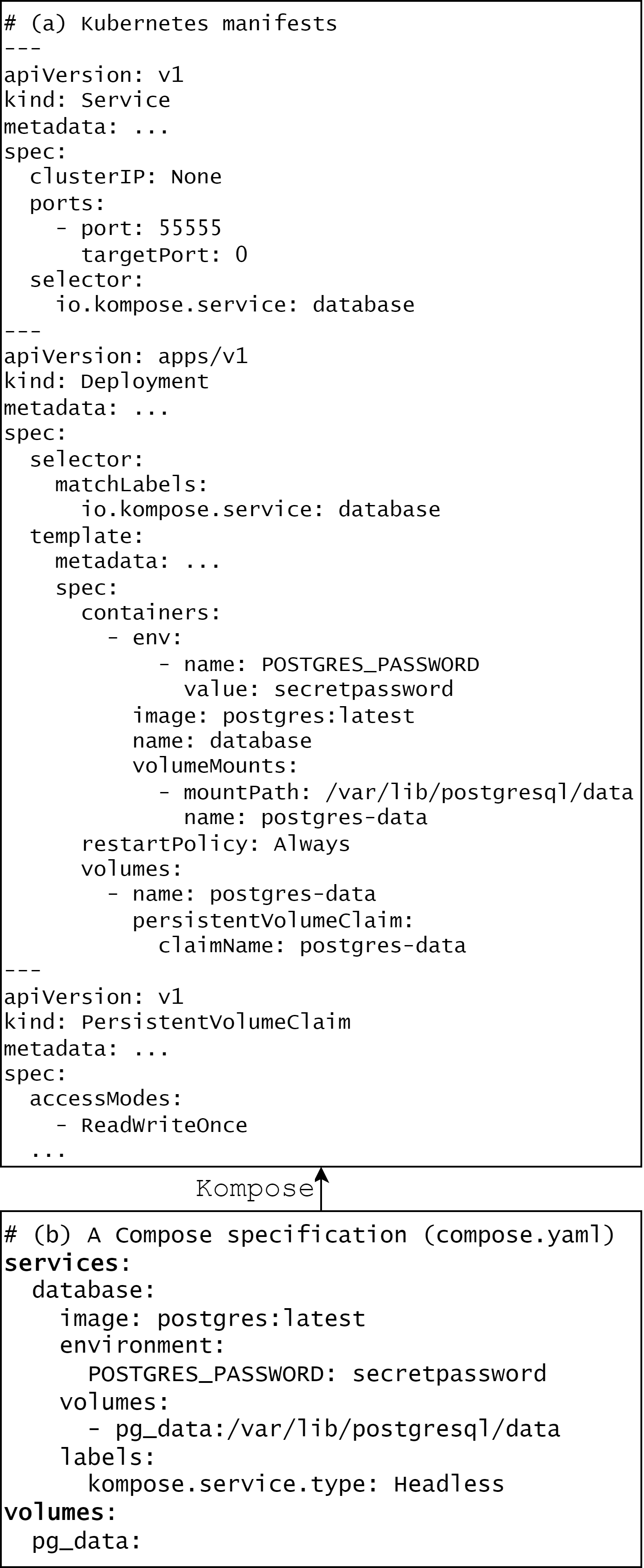}
	\caption{(a) Example of Kubernetes manifests. (b) Example of Compose specifications, including Kompose-specific labels for controlling output manifests.}
	\label{kompose-in-out}
\end{figure}

Kompose allows customization of the generated manifests by allowing users to add Kompose-specific labels to the Compose specifications.
However, to effectively utilize these features, a comprehensive knowledge of Kubernetes is essential.
As illustrated in \figref{kompose-in-out}, specifying a label directly influences the generated Service object,
causing the \verb|.spec.clusterIP| field to be set to \verb|None|, creating a \textit{headless service}.

Furthermore, our preliminary findings revealed instances in which Kompose was unable to successfully convert Compose specifications,
as demonstrated in Section \ref{sec:metrics}.

\subsubsection{Using Language Models}
Recently, LLMs, such as ChatGPT, have been widely adopted in various fields of software development.
It is now feasible to synthesize manifests by providing the LLM with a Compose specification.
This advancement has the potential to significantly streamline container deployment processes.

Based on preliminary investigations, we identified the following characteristics of the LLM-generated manifests.
First, although LLM-generated manifests are often human-readable, they frequently do not adhere to YAML specifications.
These manifests generally include placeholders, suggesting the need for manual editing.
Second, the temperature hyperparameter, which controls the randomness of the LLM outputs,
can be set to a value greater than zero to promote more creative and informative outputs.
In some cases, we obtained outputs that impressed the Kubernetes experts.
However, setting a higher temperature increases the risk of hallucinations.
This trade-off requires careful tuning of each task.
Third, the LLM outputs are nondeterministic, meaning that they do not always produce the same result.
This is a known characteristic of LLMs, even at a temperature of zero.

\section{Related Work}

\newcommand{\etal}{\textit{et al.}}

Rais \etal{}~\cite{docker-beginner}
explored the practices and challenges faced by developers when creating specifications for Docker and Docker Compose.
They found that novice developers felt that it took time to create unfamiliar specifications,
and developers rarely used support tools instead of relying on trial and error,
suggesting a high barrier for them to accept unfamiliar tools.

\subsection{Manifest Generation Using Templates}
Draft~\cite{ms-draft} from Microsoft Azure simplifies Kubernetes application development
by taking a non-containerized application and generating the Dockerfiles, Kubernetes manifests, and other artifacts.
However, this tool is currently Azure-specific, requiring custom template development for other environments.
This study aims for a solution that is not restricted to a particular environment.

Rancher~\cite{rancher} offers graphical user interface that allow users to deploy applications without writing manifests,
but we did not use this approach because of concerns regarding its generality and development costs.

\subsection{Manifest Generation Using LLMs and Quality Evaluation}

Zhang \etal{}~\cite{zhang2024} evaluated Kubernetes manifests generated by ChatGPT using static analysis tools and found that 35.8\% of 98 manifests had lint errors.
Minna \etal{}~\cite{msr2024-helm} proposed the use of LLMs to mitigate lint errors in Helm charts.
However, these studies did not assess whether the outputs satisfied expected specifications.
Furthermore, because static analysis tools require valid YAMLs, they cannot evaluate the \textit{invalid} YAMLs that LLMs frequently produce.

Several tools~\cite{chatgpt_kubernetes_yaml_plugin, kubectl_ai} demonstrate potential for generating manifests directly from natural language descriptions when formal specifications are unavailable.
However, because this study focused on leveraging existing formal specifications, that is, Compose specifications,
these tools were beyond the scope of this study.

\subsection{Application LLMs for DevOps Tasks}
DockerizeMe~\cite{dockerizeme} synthesized Dockerfiles from Python code snippets using a language model.
Two other studies~\cite{chatgpt-ansible-icwe2023,chatgpt-ansible-blog} proposed synthesis configuration scripts for Ansible.

K8sGPT~\cite{k8sgpt} is an application that uses an LLM to operate Kubernetes in a natural language.
It automates cluster operations and troubleshooting by supplementing Kubernetes knowledge with LLMs,
and its purpose is different from that of workload manifest synthesis.

\section{Problem Statement}
LLMs are still under development, and their output quality has not been clearly evaluated.
Specifically, the evaluation methods for LLM-generated manifests in prior studies had the following unresolved problems:

\begin{enumerate}
	\item Static analysis tools cannot evaluate invalid YAMLs, frequently generated by LLMs.
	\item Merely analyzing the outputs using a static analysis tool does not guarantee alignment with the input specifications.
	\item The nondeterministic nature of LLM outputs has not been adequately considered.
\end{enumerate}

\section{Method}

To address abovementioned problems, this study proposed the following methods for assessing the quality of LLM-generated manifests:
\begin{enumerate}
	\item Development of a microbenchmark suite capable of handling invalid YAMLs to evaluate their semantic correctness (\ref{sec:correctness}).
	\item Assessment of the extent to which they are grounded in the input using LLMs (\ref{sec:groundedness}).
	\item Statistical evaluation of their robustness (\ref{sec:consistency}).
\end{enumerate}

The following section provides a detailed explanation of the three quality criteria for LLM-generated manifests.

The Python code and notebooks used in this study are publicly available at \url{https://github.com/m-ueno/compose2kube}.

\begin{figure}
	\centering
	\includegraphics[width=.88\columnwidth]{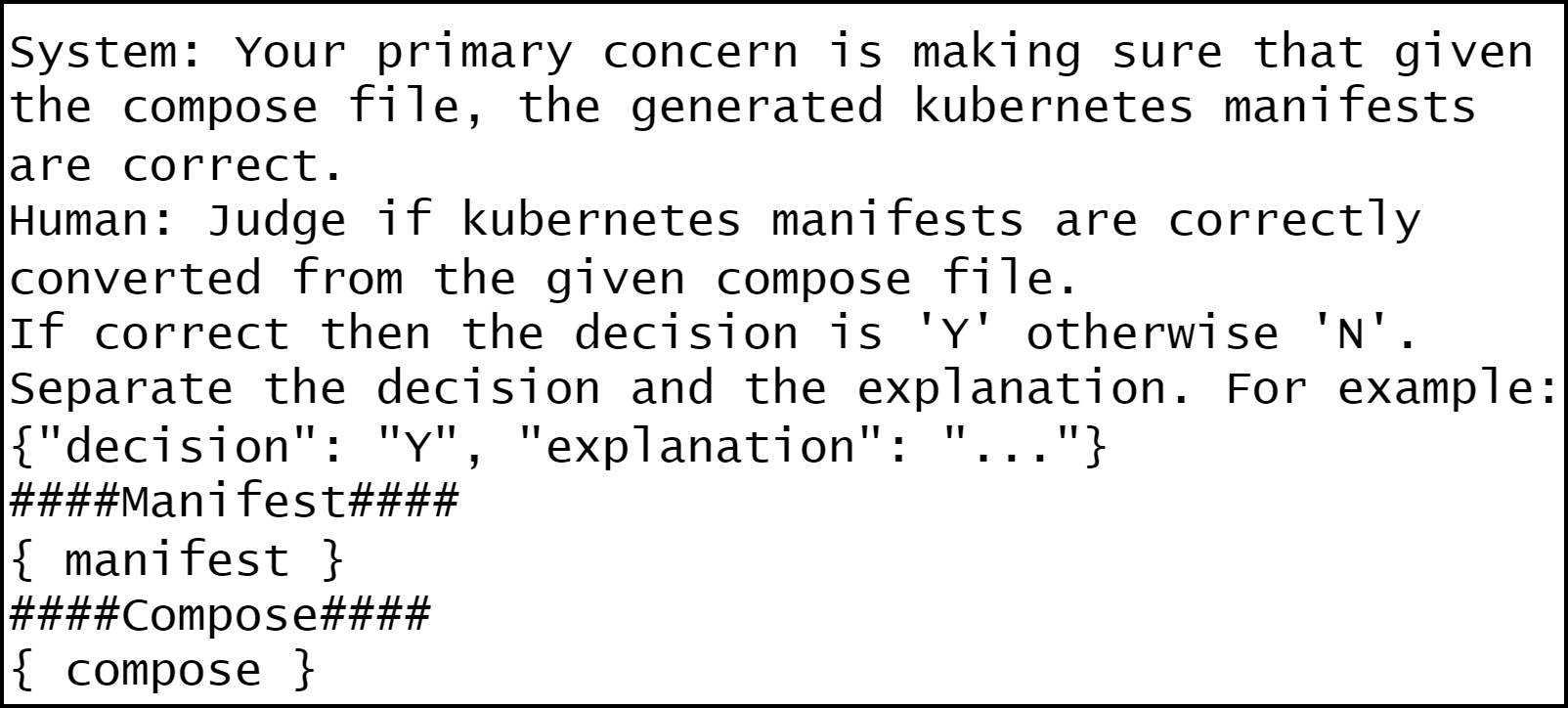}
	\caption{Prompt for evaluation of context-groundedness.} \label{fig:grader-prompt}
\end{figure}

\subsection{Correctness} \label{sec:metrics} \label{sec:correctness}
Correctness indicates that the actual output matches the expected output.
This is a binary metric (either matching or nonmatching), which is measured using a custom microbenchmark
consisting of regular expression rules and YAML-tree partial match rules~\footnote{\url{https://github.com/m-ueno/compose2kube/blob/main/src/compose2kube/benchmark/grader/rule.py}}.

The reason correctness is defined using partial match rules, rather than exact matches with the expected output,
is to evaluate manifests that include complex settings, where unique determination of the correct answers is difficult.
However, because we did not evaluate the entire manifest, defects outside the check range might have been overlooked.

We designated five microbenchmarks to evaluate correctness, each with specified inputs and expected outputs:

\newcommand{\bridging}{Bridging the gap}
\begin{enumerate}
	\item \bridging: should predict PodController
	      \begin{itemize}
		      \item Input: Include a \verb|nginx| and \verb|postgres| services.
		      \item Expected Output: The PodController for \verb|postgres| is a \verb|StatefulSet|.
		      \item Kompose Output: A \verb|Deployment|, if not specified with a label.
	      \end{itemize}
	\item Maintainability: should keep variables
	      \begin{itemize}
		      \item Input: Include environment variables.
		      \item Expected Output: Variables are output as placeholders.
		      \item Kompose Output: Variables are substituted to hard-coded string.
	      \end{itemize}
	\item \bridging: should fix invalid names
	      \begin{itemize}
		      \item Input: Include a service with an underscore (\_) in the name.
		      \item Expected Output: The underscore is replaced to satisfy Kubernetes naming rules.
		      \item Kompose Output: A Service with an underscore in its name, violating Kubernetes naming rules and result in runtime errors.
	      \end{itemize}
	\item Maintainability: should convert comments
	      \begin{itemize}
		      \item Input: Include inline comments.
		      \item Expected Output: Comments should be converted.
		      \item Kompose Output: Comments are removed.
	      \end{itemize}
	\item \bridging: should predict health check method
	      \begin{itemize}
		      \item Input: Include a \verb|health_check| field that executes a \verb|curl| command within a container.
		      \item Expected Output: An \verb|httpGet| probe.
		      \item Kompose Output: An \verb|execCmd| probe that executes a curl command within a container.
	      \end{itemize}
\end{enumerate}

\subsection{Context-groundedness} \label{sec:groundedness}
Context-groundedness represents the extent to which an output is rooted in a given context.
In other words, even if the output is valid when viewed alone, it is considered to have no basis if it does not follow the given context.
This binary metric (matching or nonmatching) was measured using the evaluation prompt shown in \figref{fig:grader-prompt}.

\subsection{Consistency} \label{sec:consistency}
Consistency is a metric that represents the variation in the output.
A smaller variation is desirable because it reflects a stable output through automatic conversion.
This metric was measured using the distribution of the number of output lines, including the comment lines.

\section{Evaluation}
\subsection{Setup}
\subsubsection{Prompts} \label{sec:eval:prompt}
Prompts are input queries or instructions provided to a language model.
We examined the following prompts to investigate their impacts on the outputs:

\begin{figure*}[!ht]
	\centering
	\includegraphics[width=.825\textwidth]{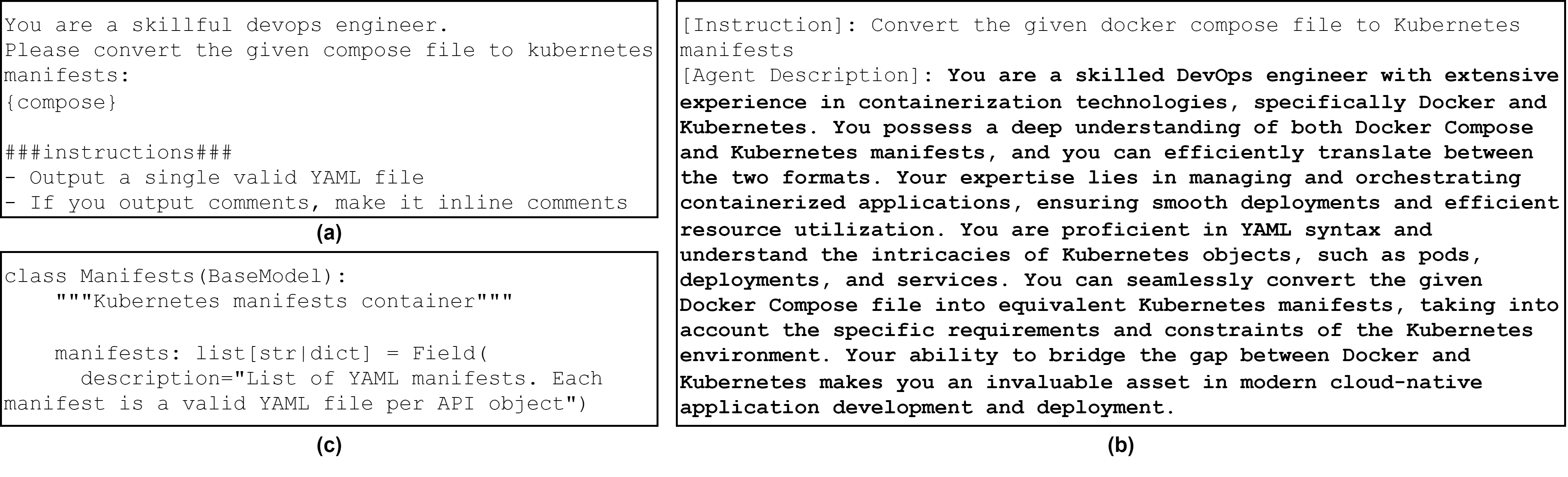}
	\caption{Prompts used to investigate impacts on output: (a) zero-shot prompts (b) role prompts extended by \expertprompting (\textbf{bold part}), and (c) schema for constraining output.
		We allowed the specification of a list of either strings or dictionaries because even if a list of strings is specified, a list of dictionaries may be returned, leading to deserialization failure.
	}
	\label{fig:prompts}
\end{figure*}

\paragraph{Zero-shot prompting}
zero-shot prompts do not include examples.
In this study, we used the simple prompt shown in \figref{fig:prompts}(a) as the baseline.

\paragraph{\expertprompting}
\expertprompting~\cite{xu2023expertprompting} is an augmented prompting technique that generates detailed expert identities for LLMs, enabling them to respond as experts and thereby enhancing their ability to provide more accurate and expert-like responses.
An example of a prompt extended by providing a zero-shot prompt is shown in \figref{fig:prompts}(b).

\paragraph{JSON mode}

JSON mode (or function calling) is a feature of the OpenAI API that outputs structured objects according to the given schema.
A schema used in this study is shown in \figref{fig:prompts}(c).

\subsubsection{Language models}

Four OpenAI models were used: GPT-3.5 Turbo, GPT-4, GPT-4 Turbo, and GPT-4o, with the temperature set to 0.7, seed set to 1, and n set to 50.

\subsubsection{Variation generation}
The prompt variations were four combinations of either normal (text) mode or JSON mode and zero-shot prompting or expert prompting.
We generated 50 manifests for each setup and subsequently aggregated the results.

\subsection{Results} \label{sec:eval:result}

\subsubsection{Consistency}
\figref{fig:eval:consistency} shows the evaluation results for consistency.
Generally, the GPT-4 and GPT-4 Turbo models exhibited less variation,
and prompts tended to have less variation in the JSON mode than in the text mode.

\subsubsection{Context-groundedness}
\figref{fig:eval:grounded} shows the evaluation results for context-groundedness, which generally indicated good performance for GPT-4 and GPT-4 Turbo.
In addition, there was little difference in the inputs owing to the difference in prompts.

\subsubsection{Correctness}
\figref{fig:eval:correctness} shows the evaluation results for correctness.
The benchmark items  (see Section \ref{sec:metrics}) with a trend for high accuracy were (2) should keep variables, (3) should fix invalid names, (5) health checks.
The benchmark items with low accuracy were (1) should predict PodController, (4) should keep comments.

In terms of models, the models achieved good results in the following ascending order: GPT-4o, GPT-4 Turbo, GPT-4, and GPT-3.5 Turbo.
In terms of prompts, we observed superior overall performance in the JSON mode compared to text mode.
In addition, ExpertPrompting generally achieved comparable or better performance than zero-shot prompting, although some results were inferior.

For benchmark (3), the (zero-shot prompt, JSON mode) pair exhibited low accuracy because the Service object was not the output.
The results showed that the performance varied substantially depending on the combination of the model and prompt.

\newcommand*{\graphwidth}{0.9\textwidth}
\begin{figure*}
	\centering
	\includegraphics[width=\graphwidth]{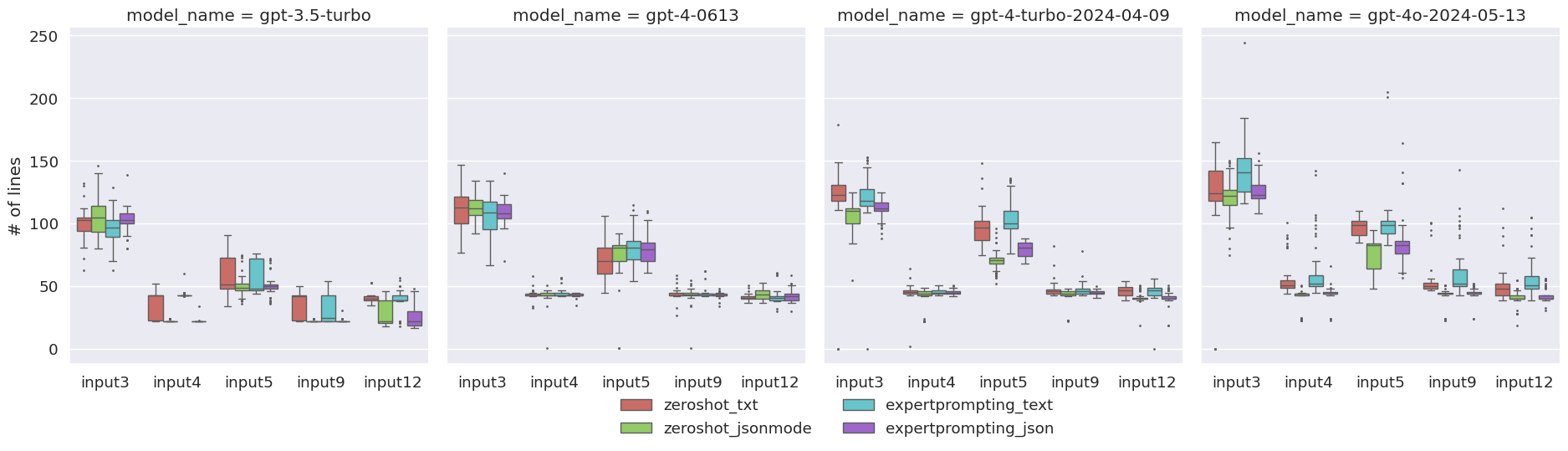}
	\caption{Evaluation of consistency. The horizontal axis represents the five types of input, the vertical axis represents the distribution of the number of lines of the output manifest, and the color difference represents the variation of prompts.}
	\label{fig:eval:consistency}
\end{figure*}

\vspace*{0cm}
\begin{figure*}
	\centering
	\includegraphics[width=\graphwidth]{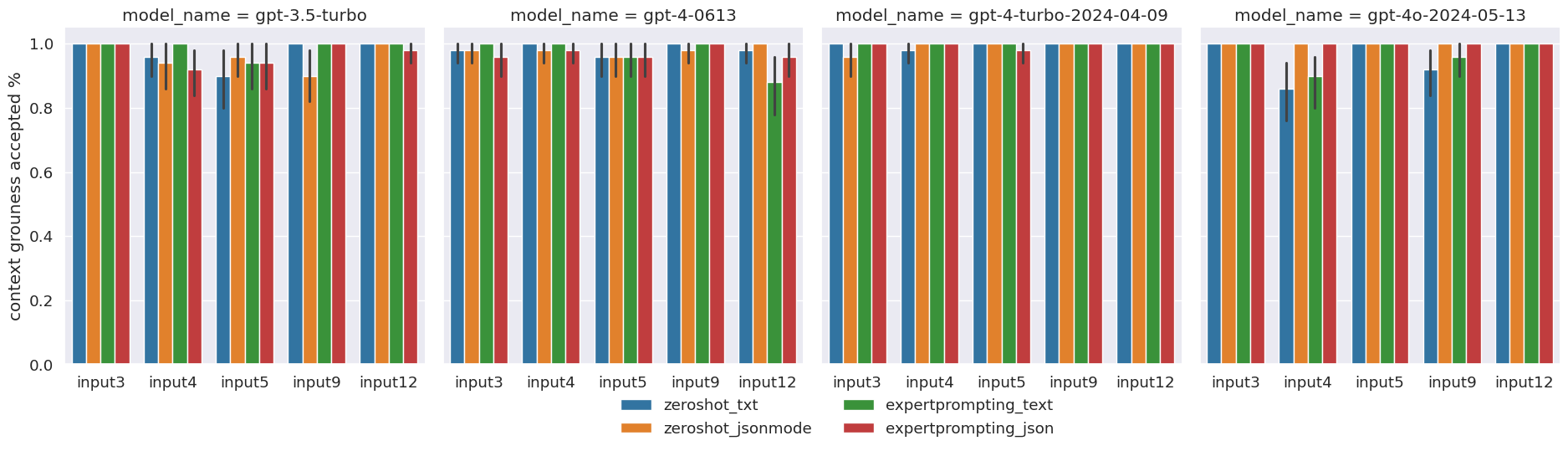}
	\caption{Evaluation of context-groundedness. The horizontal axis represents the five types of input, the vertical axis represents the success rate of the evaluation prompts, and the color difference represents the variation of prompts.}
	\label{fig:eval:grounded}

\end{figure*}
\vspace*{0cm}
\begin{figure*}
	\centering
	\includegraphics[width=\graphwidth]{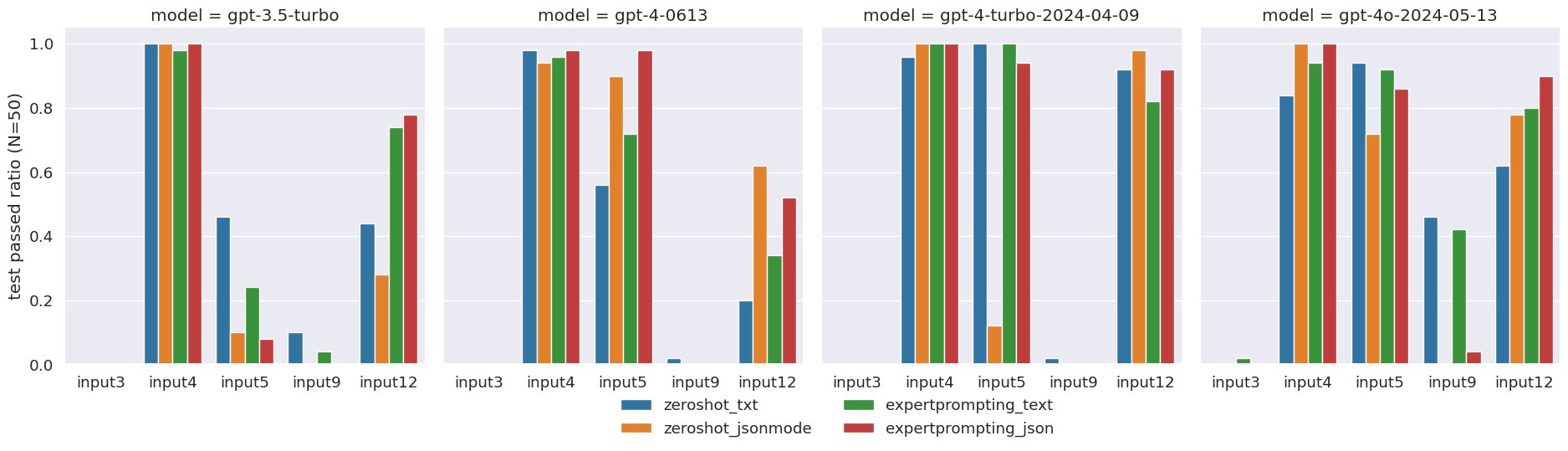}
	\caption{Evaluation of correctness. The horizontal axis represents the five microbenchmark items described in Section \ref{sec:metrics}, the vertical axis represents the success rate of the benchmark, and the color difference represents the variation of prompts.}
	\label{fig:eval:correctness}
\end{figure*}

\section{Discussion}
LLM-generated manifests often require human quality assurance owing to the probabilistic nature of their output and the challenge of eliminating hallucinations.
Although metrics such as consistency and context-groundedness can mitigate these issues, human expertise remains essential. Generating rationales, alongside generated manifests, can aid the review process.

We observed that the LLM performance varied based on the atypicality of the inputs.
For instance, non-standard use cases, such as using Docker's volume pre-population feature to implicitly share files between containers~\footnote{\url{https://github.com/m-ueno/compose2kube/blob/main/dataset/deployments_anonymized/ai-rule-editor/compose.yaml}}, pose challenges for both LLMs and human experts.

Future research should explore the quantitative evaluations of both inputs and outputs to identify patterns.
Additionally, providing users with informative feedback on the input quality, such as flagging atypical or potentially problematic aspects, can help mitigate these issues.

\section{Conclusion}
In this study, we focused on the synthesis of manifests using LLMs to develop a support technology that simplifies the process of creating Kubernetes manifests,
enabling developers to easily and correctly deploy applications in Kubernetes environments.
Thus, we developed an evaluation method based on the characteristics of the LLM output, and conducted a performance evaluation, and clarified the trend of the LLM output.

Future research directions include improving the generallty of benchmarks through the dynamic analysis of manifests,
formulating prompt tuning and pre/postprocessing pipelines for performance improvement,
and developing mechanisms to provide useful feedback to users to improve input quality.

\clearpage
\bibliographystyle{IEEEtran}
\clearpage
\bibliography{index.bib}
\clearpage
\end{document}